\documentstyle{ioplppt}

\begin{document}

\title{Gel transitions in colloidal suspensions
}

\author{J. Bergenholtz\dag\ftnote{3}{To whom correspondence should 
be addressed} 
and M. Fuchs\ddag 
}

\address{\dag\ 
Department of Physical Chemistry, G\"oteborg University, 412 96 G\"oteborg, 
Sweden
}

\address{\ddag\
Physik-Department, Technische Universit\"at M\"unchen, 85747 Garching,
Germany
}

\begin{abstract}

The idealized mode coupling theory (MCT) is applied to colloidal systems
interacting via short--range attractive interactions of Yukawa form.
At low temperatures MCT predicts a slowing down of the local dynamics
and ergodicity breaking transitions.  
The nonergodicity transitions share many features with the colloidal
gel transition, and are proposed to be the source of gelation in
colloidal systems. Previous calculations of the phase diagram are 
complemented with additional data for shorter ranges of the attractive 
interaction, showing that the path of the nonergodicity transition line 
is then unimpeded by the gas--liquid critical curve at low temperatures. 
Particular attention is
given to the critical nonergodicity parameters, motivated by recent
experimental measurements. An asymptotic model is developed, 
valid for dilute systems of spheres interacting via 
strong short--range attractions, and is shown to capture 
all aspects of the low temperature MCT nonergodicity transitions.

\end{abstract}

\maketitle

\section{Introduction}

Colloidal particles interacting with strong attractions aggregate to form
interesting structures. One such instance is gel formation, which occurs when 
the interparticle attraction is strong and of short range relative 
to the particle size. Gelation of colloidal systems appears to be a common,
if not universal, phenomenon as it has been observed experimentally
in numerous quite different colloidal--like systems. 
Notable examples of gel forming systems include 
colloid--polymer mixtures \cite{Pat89,Emm90,Poo93,Ile95,Poo95,Ver97,Poo97}, 
in which the polymer is non--adsorbing and of 
low molecular weight, and suspensions of sterically stabilized particles
in marginal solvents \cite{Jan86,Che91,Gra93,Ver95,Rue97,Rue98}. 
Emulsions \cite{Bib92}, emulsion--polymer mixtures \cite{Mel99}, 
colloid--surfactant mixtures \cite{Pia94,Kli96}, and
globular protein systems \cite{Geo94,Ros95,Mus97,Vel98} are other examples 
of colloidal systems that exhibit gel formation. 

The phase diagrams of the colloid--polymer mixtures 
have been examined in detail, revealing that they are of the 
gas--solid type without a triple point and without a liquid phase when the
attraction is short ranged. 
The disappearance of the liquid phase is a well--documented effect
\cite{Gas83,Gas86,Can89,Lek92,Tej94,Hag94,Lek95}, which is caused by the 
restricted range of 
the attraction. This situation is rather unique to colloidal systems in 
that most molecular attractions are  
of comparable range to the molecular 
dimension, precluding such phase behavior.

Whereas the equilibrium phase behavior of these systems is well understood, a
fundamental understanding of the gel transition has been more difficult to   
achieve. Colloidal gels are characterized by ramified structures 
with particles located predominantly in clusters \cite{Ver97,Poo97}. 
It is thus natural to 
attempt to describe this phenomenon with percolation theories. This may seem 
especially appropriate in that density fluctuations are very slow 
near the gel transition, and particle clusters behave 
as nearly static objects \cite{Ver95}. However, rather poor agreement 
results when the percolation transition is compared to the 
experimental results for the gel transition line in the phase diagram,
showing that the density dependence of the percolation transition is 
too strong \cite{Gra93,Ver95}.

Another suggestion has been forwarded in which the gel transition is 
attributed to dynamic percolation within a gas--liquid binodal 
which is metastable with respect 
to gas--solid coexistence \cite{Poo97}.
Such metastable binodals have been observed also in solutions of globular 
proteins. However, gel (or precipitate) formation appears to occur in 
these systems also in regions of the phase diagram outside the 
metastable binodal \cite{Mus97}, 
suggesting that the gel transition is not only triggered by
an instability towards local density fluctuations. 
The same seems to hold for sterically stabilized suspensions as well, which 
exhibit a gel transition at supercritical temperatures \cite{Ver95}.
Nevertheless, the same suspensions do form gels also under metastable 
conditions \cite{Ver95}, as do colloid--polymer mixtures 
\cite{Poo93,Ile95,Poo95,Poo97,Poo97b}, prompting careful studies of 
any connection between the gel transition 
and metastability \cite{Poo97,Poo97b}. 

We have recently suggested an explanation for the gel transition
in Ref. \cite{Ber99}, referred to henceforth as I.  
In this scenario colloidal systems form gels as a result of an arrested 
structural relaxation due to the self--trapping mechanism called the 
cage effect \cite{Got91,Got91b,Got92,Dol98}, which is the same 
effect often thought to be  
responsible for the the liquid--glass transition. The idealized mode
coupling theory (MCT), aimed at describing the cage effect in dense liquids, 
was seen to contain a bifurcation separating ergodic from nonergodic
motion also for systems of particles interacting via strong short--range 
attractive interactions. In I we attributed the nonergodic states to gel
formation; hence, our suggestion is that the gel transition can be described
within the same theoretical framework as the liquid--glass transition.  

Many of the qualitative aspects of the gel transition were found to be 
reproduced by the MCT calculations for the hard core attractive 
Yukawa system (HCAY).
The calculated phase diagrams were found to exhibit gel transition lines that
connect smoothly with the hard sphere glass transition at high temperature
and extend to the critical and subcritical regions at low temperature along
paths that depend
critically on the range of the attraction. The phase diagram obtained by
Verduin and Dhont \cite{Ver95} 
is qualitatively reproduced by the MCT when the attraction
is of intermediate range such that the gel transition meets the critical point.
For shorter--range attractions the gel transition passes above the critical
point, suggesting that structural arrest occurs instead of  
gas--liquid phase separation, which appears to agree with some measurements
on sterically stablized particle systems \cite{Che91,Gra93,Rue97,Rue98}.

A finite zero--frequency shear modulus is predicted by MCT, 
in agreement with many
measurements on colloidal gels \cite{Poo97,Che91,Gra93,Rue98}, 
which is expected to be intimately connected
to a finite yield stress observed in other measurements \cite{Ver96}. 
At present the theory cannot account for the density dependence of
the modulus , indicating that the mesoscopic structure of the gels
is important and is not captured correctly. In addition, the theory
alone cannot explain the growth of the small angle 
peak of the static structure factor nor the
fractal scaling of its peak position observed on quenching of many 
suspensions \cite{Poo97}. It does, however, provide an explanation for why
the growth of the small angle peak slows down and arrests following 
deeper quenches. Very differently from earlier approaches, which
attributed the ultimate gel arrest to a global jamming or percolation
transition, our microscopic theory predicts the structural arrest
to be driven by anomalies of and a slowing down of the local dynamics.  

The purpose of this article is to provide a more detailed account
of the results of the model study in I. In particular, the nonergodicity 
parameters (cf. Section \ref{mct}) for several attraction ranges 
will be shown, motivated by the recent measurements on colloid--polymer systems 
by Poon and co--workers \cite{Poo99}. 
In addition, a detailed description is given of 
the asymptotic model developed in I. This model captures the 
relevant features associated with our suggested scenario for gel formation. 
A discussion of the relevance of these results is also included.  

\section{Theory of colloidal gelation}

\subsection{Mode coupling theory}
\label{mct}

The idealized mode coupling theory (MCT) assumes that the dominant 
mechanism for structural relaxation in dense liquids is the cage effect.
At short times particles are trapped in the surrounding cage of 
neighboring particles. At longer times particle escape from the cage
leads to structural relaxation to equilibrium. For sufficiently strong
interactions a bifurcation occurs in the governing equations --- interpreted as 
the permanent entrapment of particles within
their cages --- causing the intermediate scattering function $F_q(t)$ to 
acquire a non--zero long--time limit known as the nonergodicity parameter,
glass form factor, or Debye--Waller factor $f_q = F_q(t\to \infty )/S_q$,
where $S_q$ is the structure factor and $q$ is the modulus of the wavevector. 
The transition from a vanishing to a finite long--time limit of $F_q(t)$
is discontinuous and defines the liquid--glass transition within MCT.
It is not a conventional thermodynamic phase transition, but rather an
entirely dynamic transition, interpreted generally as an ergodic--nonergodic
transition. This simplified scenario captures many aspects of the
liquid--glass transition in molecular liquids \cite{Got91,Got92} as well as 
hard sphere suspensions \cite{Got91b,Got92,Pus87,vMe91,vMe94,vMe95}.  

The governing MCT equation for the time evolution of the intermediate scattering
function reduces in the long--time limit to  
\begin{eqnarray}
\frac{f_q}{1-f_q} &=& \frac{\rho }{2(2\pi )^3 q^2}\int d{\bf k}
V({\bf q},{\bf k})^2 S_qS_kS_{|{\bf q}-{\bf k}|}f_kf_{|{\bf q}-{\bf k}|}
\nonumber \\
V({\bf q},{\bf k}) &=& {\hat {\bf q}}\cdot ({\bf q}-{\bf k})
\; c_{|{\bf q}-{\bf k}|}  - {\hat {\bf q}}\cdot {\bf k} \; c_k
\label{coh}
\end{eqnarray}
where $\rho $ is the number density and $c_q=(1-S_q^{-1})/\rho $,
which appears in the vertex function $V({\bf q},{\bf k})$, is the
Fourier--transformed direct correlation function. 
Consideration of the single particle motion leads to another set of
equations for the incoherent nonergodicity parameter $f_q^s$ (also known as
the Lamb--M\"ossbauer factor)
\begin{eqnarray}
\frac{f_q^s}{1-f_q^s} &=& \frac{\rho }{(2\pi )^3 q^2}\int d{\bf k} 
V^s({\bf q},{\bf k})^2 S_kf_kf^s_{|{\bf q}-{\bf k}|} \nonumber \\
V^s({\bf q},{\bf k}) &=& {\hat {\bf q}}\cdot {\bf k} \; c_k
\label{incoh}
\end{eqnarray}  
where $f_q^s=F_q^s(t\to \infty )$, with the self--intermediate scattering
function $F_q^s(t)$. 

For a given $S_q$, equations \ref{coh} and \ref{incoh} are 
closed equations for $f_q$ and $f_q^s$.
They are solved numerically by iteration starting from the initial iterate 
$f_q=f_q^s=1$. With this starting point the iteration converges monotonically
to the correct solution for the nonergodicity parameters, given by the
largest solution to the equations \cite{Got95}. The wavevector integrations are
performed efficiently using Simpson's rule on a uniformly discretized grid:
$q= i\Delta q$, $i=0, \ldots , N$. The critical glass transition boundary
is identified by bracketing of the given input conditions, such as 
temperature and density, which delineates regions where $f_q$ is zero and
finite; regions of the phase diagram which result in finite nonergodicity 
parameters are identified as glass states. Most calculations were done 
with the parameters $\Delta q=0.3\sigma ^{-1}$ and $N=800$. 

\subsection{Model system}
\label{HCAYsec} 

The hard core attractive Yukawa (HCAY) interaction potential captures both 
short--range excluded volume interactions and 
variable--range attractive particle interactions.
It is given by 
\begin{equation}
u(r)/k_BT = \left\{ \begin{array}{ll}
         \infty & 0 < r < \sigma \\
 -\frac{ K}{r/\sigma }e^{-b(r/\sigma -1)}   
\;\;\; &  \sigma < r
\end{array}
\label{hcay}
\right.
\end{equation}
where the dimensionless parameter $K$ regulates the depth of the attractive
well and the reduced screening parameter $b$ sets the range of the attraction.

The equilibrium phase diagrams for a number of attraction ranges
have been obtained by computer simulations \cite{Hag94}, showing indeed that 
the liquid phase disappears upon restricting the range of the
attraction. Because the HCAY static structure factor is known 
semi--analytically from the mean spherical approximation (MSA) 
\cite{Wai73,Hoy77,Cum79}, it
is convenient to base our study on this case. The  penalty is that
the resulting predictions are limited not only by the approximations
made in the MCT, but also those made in the MSA.
Therefore, the results should be viewed as qualitative rather than quantitative.
  
\subsection{Asymptotic model}
\label{Asymsec}

In the following we describe an asymptotic model, originally developed in I, 
that captures the relevant
features of the suggested gelation mechanism. 
At low densities ($\phi = \pi \rho \sigma ^3 /6 \to 0$) 
the Ornstein--Zernike direct correlation 
function becomes independent of density. Specifying this to the MSA of the HCAY
fluid, we obtain $c_q \to Kb(\sigma /b)^3{\tilde c}(q,\sigma ,b)$, where
${\tilde c}(q,\sigma ,b)=(b/\sigma )^2\int_{r>\sigma }d{\bf r}e^{-b(r/\sigma -1)}
e^{i{\bf q}\cdot {\bf r}}/r$ in which we have neglected the contribution from
the hard core which is independent of $K$ and $b$ at low densities. 
In the limit of strong attractive interactions at low densities, 
the following scaling simplifies the MCT equations
\begin{equation}
\phi \to 0 \;\;\;\;\; \mbox{and}\;\;\;\;\; K\to \infty , \; \mbox{so that}
\;\;\;\; \Gamma = \frac{K^2\phi }{b}= \mbox{constant}
\label{asymp}
\end{equation}
In addition, $S_q\to 1$ in this limit and the MCT equations 
(\ref{coh} and \ref{incoh}) simplify considerably such that 
the nonergodicity transitions occur at 
$\Gamma = \Gamma_c(b)$, leading to the asymptotic prediction 
$K_c \propto 1/\sqrt{\phi }$. 

For short--range attractive interactions, in the limit $b\to \infty $, a further
simplification arises because the MCT vertex functions become linear functions
of $\Gamma $ only, if we introduce the rescaled wavevectors ${\tilde q}=q\sigma /b$.
Equation \ref{coh} simplifies to
\begin{equation}
\frac{{\tilde f}_{\tilde q}}{1-{\tilde f}_{\tilde q}}=
\frac{\Gamma }{{\tilde q}^2}\int d{\tilde {\bf k}}
{\tilde V}({\tilde {\bf q}},{\tilde {\bf k}})^2
{\tilde f}_{\tilde k} {\tilde f}_{|{\tilde {\bf q}}-{\tilde {\bf k}}|}
\label{coh-resc}
\end{equation}
with the dominant contribution to the rescaled vertex function as
\begin{eqnarray}
{\tilde V}({\tilde {\bf q}},{\tilde {\bf k}})^2 &=&
{\tilde V}^s({\tilde {\bf q}},{\tilde {\bf q}}-{\tilde {\bf k}})^2
+{\tilde V}^s({\tilde {\bf q}},{\tilde {\bf k}})^2
\nonumber \\
{\tilde V}^s({\tilde {\bf q}},{\tilde {\bf k}})^2 &=& 
\frac{3}{\pi ^2 } \frac{({\tilde {\bf q}}\cdot {\tilde {\bf k}})^2}{
{\tilde q}^2{\tilde k}^2(1+{\tilde k}^2)}
\label{vertex-resc}
\end{eqnarray}
where the nonergodicity parameters depend only on the rescaled wavevectors
$f_q \to {\tilde f}_{\tilde q}$.
The same set of equations is found to govern the incoherent 
nonergodicity parameters ${\tilde f}^s_{\tilde q}$. 
Note that in this limit single particle and collective nonergodicity 
factors are identical and exhibit the small wavevector expansion 
${\tilde f}_{\tilde q}=1-{\tilde q}^2\left(r_sb/\sigma \right)^2$, 
where $r_s$ is the localization length, or root--mean--square displacement, 
in the glass.

In this limit the coupling constant $\Gamma $ assumes a 
unique value at the transitions, which is found from solving 
equations \ref{coh-resc}
and \ref{vertex-resc} numerically resulting in $\Gamma _c \approx 3.02$ for
$b\to \infty $. We emphasize that these asymptotic forms are valid for the HCAY
systems using the MSA in the prescribed limits. 

\section{Results}
\label{Ressec}

\subsection{Phase diagrams}

Using the MSA static structure factor \cite{Wai73,Hoy77,Cum79} as input,
the MCT was solved for several screening parameters: $b$ = 7.5,
20, 30, and 40. The progression of the glass transition can be traced from the
(Percus--Yevick) hard sphere limit, corresponding to K=0,
to lower temperatures in terms of the reduced temperature K$^{-1}$.
In I we showed the phase diagrams for $b$ = 7.5, 20, and 30. 
This work adds an additional phase diagram for $b=40$. 

As shown in I, at low temperatures (large values of $K$) 
the glass transition is traced along
different paths in the phase diagrams depending on the value of the 
screening parameter, i.e. the range of the attractive interaction.
In all cases they bend towards lower densities when the 
temperature is decreased sufficiently. 
The transition lines for 
intermediate--range attractions ($b$ = 7.5 and 20) reach subcritical 
temperatures on the liquid side of the spinodal.
For shorter ranges of the attraction ($b=30$) the nonergodicity
transition line lies entirely within the fluid phase 
above the two phase region, and
extends to subcritical temperatures at 
low densities on the vapor side of the spinodal. 
This is shown in detail in \fref{PH3}, in which the results
for $b$ = 30 and 40 are shown. As seen, decreasing the attraction range 
further to $b=40$ causes the nonergodicity transition line to move away 
from the spinodal curve. 
In contrast to the MSA phase diagrams studied in I, the $b$ = 40 
transition line is located sufficiently far away from the spinodal
curve to remove the additional nonergodicity transition line
that appears in the MSA phase diagrams with lower $b$.
This type of transition is discussed in the Appendix of I.
Also shown in \fref{PH3} are the spinodal
curves, defined by the condition $S_q\to \infty $ for $q\to 0$.
In the present context they are shown as an indication of where
gas--liquid phase separation is likely to occur, provided a liquid phase 
is present. 

As pointed out in I, the MCT nonergodicity transition lines 
resemble the gel transition lines determined experimentally. Most notably,
the diagram with $b=20$ (see I), which displays a near meeting of the  
nonergodicity transition and the critical point, bears a strong resemblance 
to the phase diagram determined by Verduin and Dhont \cite{Ver95} 
in their measurements of sterically stabilized silica suspensions. 
Other measurements on sterically stabilized suspensions found a 
gel transition line with no apparent evidence of phase separation,
which can be expected, according to the MCT, 
for attractions of very short range, such as $b=30$ or $40$ in \fref{PH3}.

One important aspect of the MCT nonergodicity transitions is that they are
not induced by long--range structural correlations associated with critical
fluctuations. This is demonstrated by the asymptotic model which is valid
in the limit in equation \ref{asymp} and $b\to \infty $ for which
$S_q\approx 1$. 
The predictions of the model are shown in \fref{PH3} 
as chain curves. There is quantitative agreement at low densities
and the model predictions reproduce the results of the full calculation 
for the nonergodicity transitions qualitatively at moderately high densities. 
The agreement with the asymptotic model demonstrates that 
the $q\to 0$ limit of $S_q$ is decoupled from the structural 
arrest at low densities and temperatures. This conclusion is of 
particular importance as 
not all relevant mode couplings expected 
for systems near the critical point \cite{Kaw76} are included
in this version of MCT. 

\subsection{Nonergodicity parameters}

Dynamic light scattering measurements of the 
nonergodicity parameters led, inter alia, to the identification
of the glass transition for the colloidal hard sphere system 
\cite{Pus87,vMe91,vMe94,vMe95}.
As MCT predicts very different behavior for these when
strong short--range attractions are present \cite{Ber99}, it can be expected  
that such measurements will provide a similarly decisive test of the MCT 
of gelation. For that reason we focus here on the behavior of the 
nonergodicity parameters as functions of density and temperature.

When the attraction is of moderately short range, such as for $b=7.5$,
\fref{FQ7p5} shows that the coherent nonergodicity parameters 
deviate little from the hard sphere $f_q$ as the temperature is
lowered, with the exception of an increasing $q \to 0$ value.  
Note that the $f_q$ shown in \fref{FQ7p5}  
and later figures are the critical nonergodicity parameters,
which appear along the MCT transition lines shown 
in figure 2 of I and \fref{PH3}
of this work. The increase of the $q \to 0$ value results in this 
case mainly from the increased contribution of long--range correlations, 
which derive from the increase of the $q\to 0$ limit of $S_q$ 
upon approaching the liquid side of the spinodal curve.
Shown also in \fref{FQ7p5} is the prediction from the asymptotic
model based on equation \ref{asymp} and $b\to \infty $, which does not capture
the behavior of $f_q$ at the lowest temperatures. This should be expected as
the attraction is not particularly short ranged.
   
Decreasing the range of the attractive interaction to $b=20$, which causes
the nonergodicity transition line to move closer to the critical point
(see figure 2 of I), leads now to form factors that deviate more 
from hard sphere behavior at low temperatures. 
Compared to $b=7.5$, this shorter range 
attraction causes coupling among more wavevector--modes, leading to an
increase in the width of $f_q$ as the temperature is decreased. 
This effect is produced by the increased width of the static structure factor,
caused by strong short--range correlations due to the attraction. 
As in the case of $b=7.5$, the $q\to 0$ limit of $f_q$ for $b=20$ increases
when the temperature is decreased. Now, however, the asymptotic model 
yields a reasonably accurate prediction for $f_q$ at the lowest temperatures, 
which can be expected from the rather close agreement between the 
asymptotic model and the full calculation of the transition line
for $b=20$.  

When the range of the attraction is decreased further by changing 
the HCAY screening parameter to $b=30$,
the path of the nonergodicity transition is unimpeded by the gas--liquid 
critical curve
(see \fref{PH3}); it extends to subcritical temperatures on the 
low density, vapor side of the spinodal curve. The form factors
along the transition, shown in \fref{FQ30}, 
are qualitatively similar to those corresponding to $b=20$, 
except that the width of $f_q$ is further increased due to stronger 
short--range correlations. The agreement with the asymptotic model is 
improved at the lower temperatures where the oscillations in $f_q$ are 
now somewhat suppressed. The improved
agreement should be expected because the asymptotic model is based in part
on the limit $\phi \to 0$, which can be fulfilled by this system as the
nonergodicty transition line lies entirely in the (very likely metastable)
single phase fluid regime.  

The dynamics of the gel transitions can be expected to be anomalously
stretched over many orders in time and to exhibit a rapid slowing down
upon approaching the nonergodicity transition lines.
As discussed in detail in reviews of the MCT \cite{Got91,Got92},
this and a number of other qualitative aspects can be understood
from the factorization property and asymptotic expansions, which describe
the sensitive variation of the cage dynamics close to the transition. 
One finds: $F_q(t)/S_q = f^c_q + h_q G^\lambda (t)$. As the so--called
$\beta $--correlator, $ G^\lambda (t)$, also exhibits numerous
universal features depending on one material parameter only, 
the exponent parameter $\lambda $, this expression provides for
crucial experimental tests of MCT transition scenarios, such as those performed
for hard sphere colloids \cite{vMe91,vMe94,vMe95}.
We find $\lambda = 0.89$ \cite{Ber99}, predicting a very anomalous stretching
and rapid increase of the longest relaxation time \cite{Got91,Got92}.
\Fref{ampl} shows the two wavevector dependent amplitudes, which
describe the gel structure ($f^c_q$, also included in figures \ref{FQ7p5}--
\ref{FQ30}) and localized cage dynamics ($h_q$). The amplitude factor $h_q$, 
which describes the spatial extent of that dynamical process 
which arrests at the
gel transition, is found to be peaked at rather large wavevectors stressing
that the local motion of the colloids is suppressed at the gel transitions.
 
\section{Discussion and Conclusions}
\label{Discsec}

Structural arrest as described by the idealized MCT is found to be a
plausible explanation for colloidal gelation. The underlying cause of gelation
in this scenario is a breaking of ergodicity caused by strong short--range
attractions in dilute systems. An ergodic--nonergodic transition is
characteristic also for the glass transition within the framework of MCT.
The low temperature gel transitions are accompanied by the 
cessation of hydrodynamic diffusion and the appearance of rather 
large finite elastic moduli due to the particles being tightly 
localized in ramified clusters (for more details see I).

At relatively high temperatures, corresponding to weak attractions
among the suspended particles, the cage surrounding a typical particle
is distorted by dimerization \cite{Fab99} (see also \cite{Ber99}). 
The cage must be reinforced by increasing
the critical colloid density before structural arrest ensues, which
leads to a glass transition line that moves initially towards 
higher density with decreasing temperature. 
This trend appears to require hard core repulsions and becomes more
pronounced when the range of the attractive interaction decreases.   
It has been observed in the adhesive hard sphere system
\cite{Ber99,Fab99} and the Yukawa systems investigated in I and this study. 
For the colloid--polymer mixtures recrystallization 
of glassy samples has been reported upon introducing a small concentration of 
low molecular weight non--adsorbing polymer \cite{Ile95}.
We attribute this effect to the glass transition 
line possessing an initial slope in accordance with the MCT predictions, i.e. 
in the direction of higher density with decreasing temperature. 

Stronger short--range attractions cause particle aggregation, 
leading in effect to an increase of the density in the local 
environment of a typical particle. According to MCT, aggregation can lead to 
structural arrest of the long--time dynamics despite the bulk density 
being much lower than the hard sphere glass transition density.   
As shown in I and \fref{PH3}, the precise path of the nonergodicity 
transition line in the phase diagram
depends now critically on the range of the attraction, reminiscent 
of the dependence of the freezing line on the range of the 
attraction \cite{Gas83,Gas86,Can89,Lek92,Tej94,Hag94,Lek95}. 

The low density nonergodicity transitions, which appear only for 
sufficiently short ranges of the attraction, are not only caused by the 
excluded volume effect which dominates at higher densities, 
but are additionally affected
by the low-- and high--$q$ behavior of the static structure factor.
The asymptotic model demonstrates, however, that the short--range 
(high $q$) correlations become increasingly important as the range of 
attraction is  decreased, such that they are the dominant cause of the 
structural arrest at low densities. This observation indicates that the 
nonergodicity transitions are unaffected by long--range correlations 
and the presence of the critical point. 
Thus the fact that the MCT transitions occur
in close proximity to the spinodal curve for a significant range of $b$
is merely coincidental. The results shown in \fref{PH3} are in accord
with this conclusion; the nonergodicity transition line for $b=40$ is 
seen to be located further away from the critical curve than for $b=30$. 

Nevertheless, several studies, particularly the extensive ones of the 
colloid--polymer mixtures, reveal what appears to be an intimate connection
between the gel transition and the metastable gas--liquid binodal.
It is possible that the gel transition changes character once it crosses
the metastable binodal into the metastable and unstable regions. 
This is not in disagreement with our description of the arrest
of the local dynamics, but indicates that the large--distance dynamics
can be more complicated.
Unfortunately we are yet unable
to make any predictions inside the unstable region with the present
MCT because an appropriate $S_q$ is not available and the theory 
assumes closeness to equilibrium \cite{Got92}.  

Krall and Weitz \cite{Kra98} have studied low density gels 
experimentally at small wavevectors, though still
larger than the wavevectors characterizing the small angle scattering peak. 
The low density gels are 
associated with finite nonergodicity parameters, supporting the MCT gelation
mechanism. However, Krall and Weitz measure relatively small values for
the nonergodicity parameters, whereas the full MCT solutions, as well as the
asymptotic model described in Section \ref{Asymsec},
predict large values for $f_q$ at small wavevectors.
Such large values are observed experimentally also, but only at somewhat
higher densities.
We are attempting presently to locate a low density region in the
MSA phase diagrams with this type of dynamics. 


Experimental measurements of the dynamics of gels at 
higher densities have been conducted.
Sterically stabilized suspensions
\cite{Ver95} and colloid--polymer mixtures with low molecular 
weight polymer \cite{Poo99} show that these gels are associated 
also with finite 
nonergodicity parameters, further supporting the present model calculations.
Moreover, the measured $f_q$ assume values not too much below unity both 
at low $q$ \cite{Ver95} and at values near the
$q$ corresponding to the principal peak of $S_q$ \cite{Poo99}. 

Poon and co--workers \cite{Poo97b} 
have made detailed studies of the low density and 
temperature region of the colloid--polymer phase diagram for low molecular
weight polymers. They identified a variety of different dynamics which
can be reconciled with the MCT calculations provided their system belongs
to the HCAY diagram with $b\approx 20$. 
For such a situation the gel transition meets the spinodal on the liquid
side, without interfering with the critical behavior along the vapor side.
Conjecturing that the gel transition can exist in the unstable region
as found experimentally by Verduin and Dhont \cite{Ver95}, 
we obtain a phase diagram 
qualitatively similar to that determined by Poon et al. \cite{Poo97b}. 
We expect that quenching of a suspension at low to moderately low density 
results initially in normal phase separation dynamics,
characterized by either spinodal decomposition or nucleation and growth.
This process proceeds until the denser domains reach the critical 
density for gelation, subsequently arresting the structure, which would be
consistent also 
with the measurements in Ref. \cite{Ver95}. Thus we anticipate that the
so--called transient gelation region, discovered by Poon et al. \cite{Poo97b}
for colloid--polymer systems with short polymers, 
corresponds to a nonergodicity transition in the unstable region of the 
phase diagram.

The asymptotic model of Section \ref{Asymsec} provides additional 
support for this interpretation. First, arrest of long--range structures
is caused by an arrest of the local dynamics. In contrast to percolation
approaches for the gel arrest,
the local density fluctuations also exhibit slowing down and dynamical
anomalies. Second, the tight localization of the particles provides a
rationale for using concepts like bond--formation and sticking
probabilities, even though the particles obey diffusive equations of motion.
Third, the approach to unity of the collective nonergodicity or Debye--Waller
factors for small wavevectors, $f(q)\to 1$ for 
$q\to 0$, indicates that the particles are bound to infinite 
clusters in this limit. Momentum conservation, or Newton's law of 
action--and--reaction, otherwise would require $f(q\to 0) <1$, as found for
glasses. 

In summary, gelation in colloidal systems is attributed to a breaking
of ergodicity, captured qualitatively by the MCT applied to the HCAY system. 
The calculated phase diagrams show many similarities with those
determined experimentally for colloid--polymer mixtures and sterically
stabilized suspensions. Decreasing the range of the attractive interaction
sufficiently causes the gel transition to move further into the 
single phase region. The gross features
of the MCT nonergodicity parameters at low temperatures are seen to be 
in accord with the few existing measurements, supporting the proposed
gelation mechanism. 
 
\subsection{Acknowledgments}

Financial support by the 
Deutsche Forschungsgemeinschaft (DFG)
(Grant No. Fu309/2-1) is gratefully acknowledged.



\Figures

\begin{figure}
\caption{
Enhancement of the low density and temperature region of the HCAY diagrams for
$b=30$ and $40$. The MSA spinodal curves are shown with the
critical points denoted by $(\bullet )$, together with
the corresponding MCT nonergodicity transition lines as labeled.
The chain curves correspond to the asymptotic prediction
in equation \protect \ref{asymp} and $b\to \infty $
with $\Gamma _c(b\to \infty )=3.02$. The inset shows the phase diagram
for $b=40$ for the same density and temperature ranges as in figure 2 of I.
}
\label{PH3}
\end{figure}

\begin{figure}
\caption{HCAY critical coherent nonergodicity parameters $f_q$
for $b$=7.5 along the critical boundary as functions of normalized wavevector
$q\sigma $ and Yukawa prefactor $K$.
Note that the volume fraction varies
according to the $b$=7.5 nonergodicity transition
line in figure 2 of I. The $K$ values,
incremented by 1, run from bottom--to--top from 0 to 7.
The curve shown in bold
is the  asymptotic prediction resulting from equation \protect \ref{asymp}
and $b\to \infty $.
}
\label{FQ7p5}
\end{figure}

\begin{figure}
\caption{HCAY critical coherent nonergodicity parameters $f_q$
for $b=20$ as in figure \protect \ref{FQ7p5}. The $K$ values,
incremented by 2, run from bottom--to--top from 0 to 12.
The asymptotic prediction
from equation \protect \ref{asymp} is shown as the bold curve.
}
\label{FQ20}
\end{figure}

\begin{figure}
\caption{
HCAY critical coherent nonergodicity parameters $f_q$
for $b=30$ as in figure \protect \ref{FQ7p5}. The $K$ values,
incremented by 2, run from bottom--to--top from 0 to 16.
The asymptotic prediction
from equation \protect \ref{asymp} is shown as the bold curve.
}
\label{FQ30}
\end{figure}

\begin{figure}
\caption{
Critical coherent nonergodicity parameter $f_q$ and amplitude factor
$h_q$ as functions of the scaled wavevector for the asymptotic
model in equations \protect \ref{asymp} -- \protect
\ref{vertex-resc} and $b\to \infty $.
}
\label{ampl}
\end{figure}

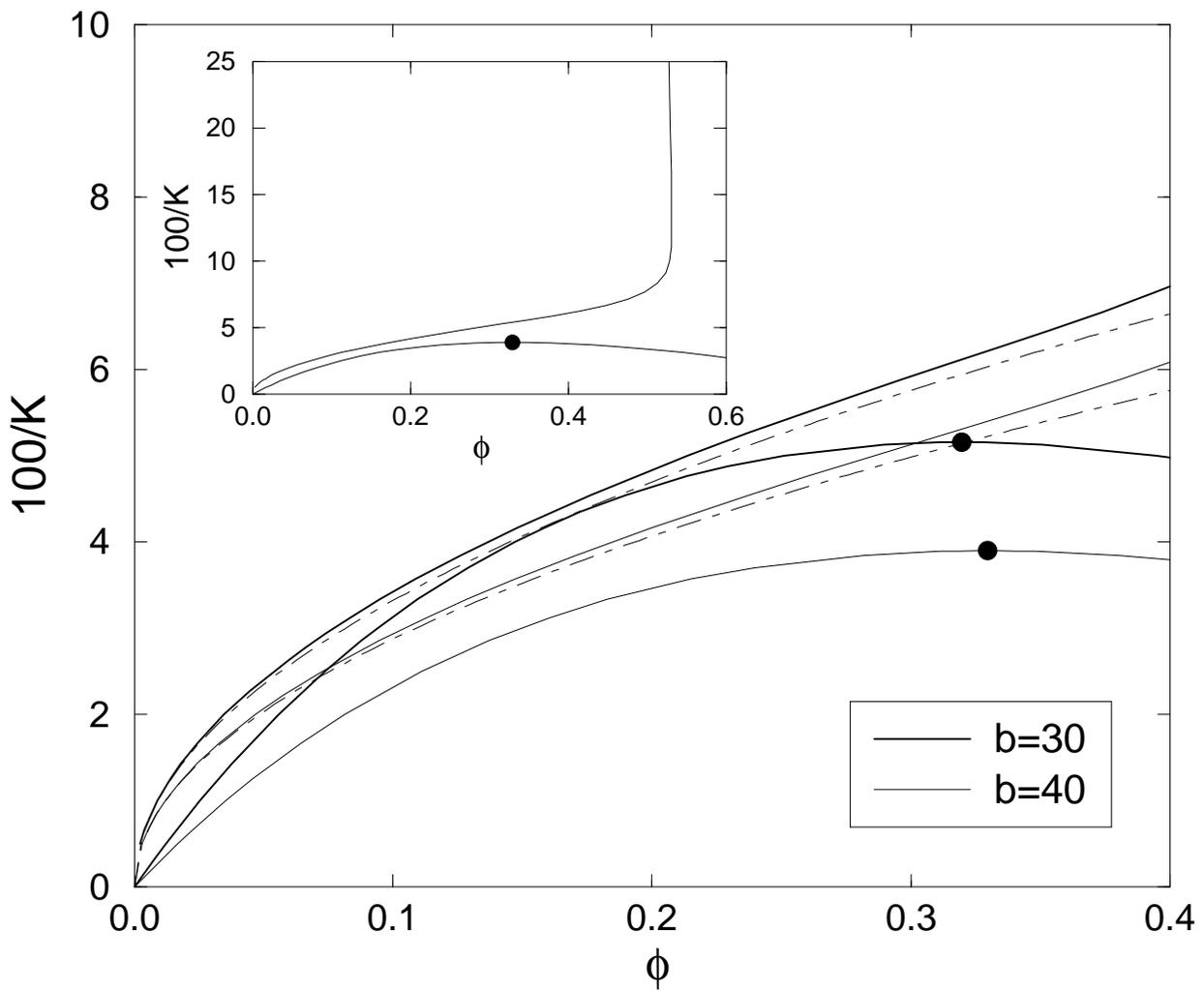
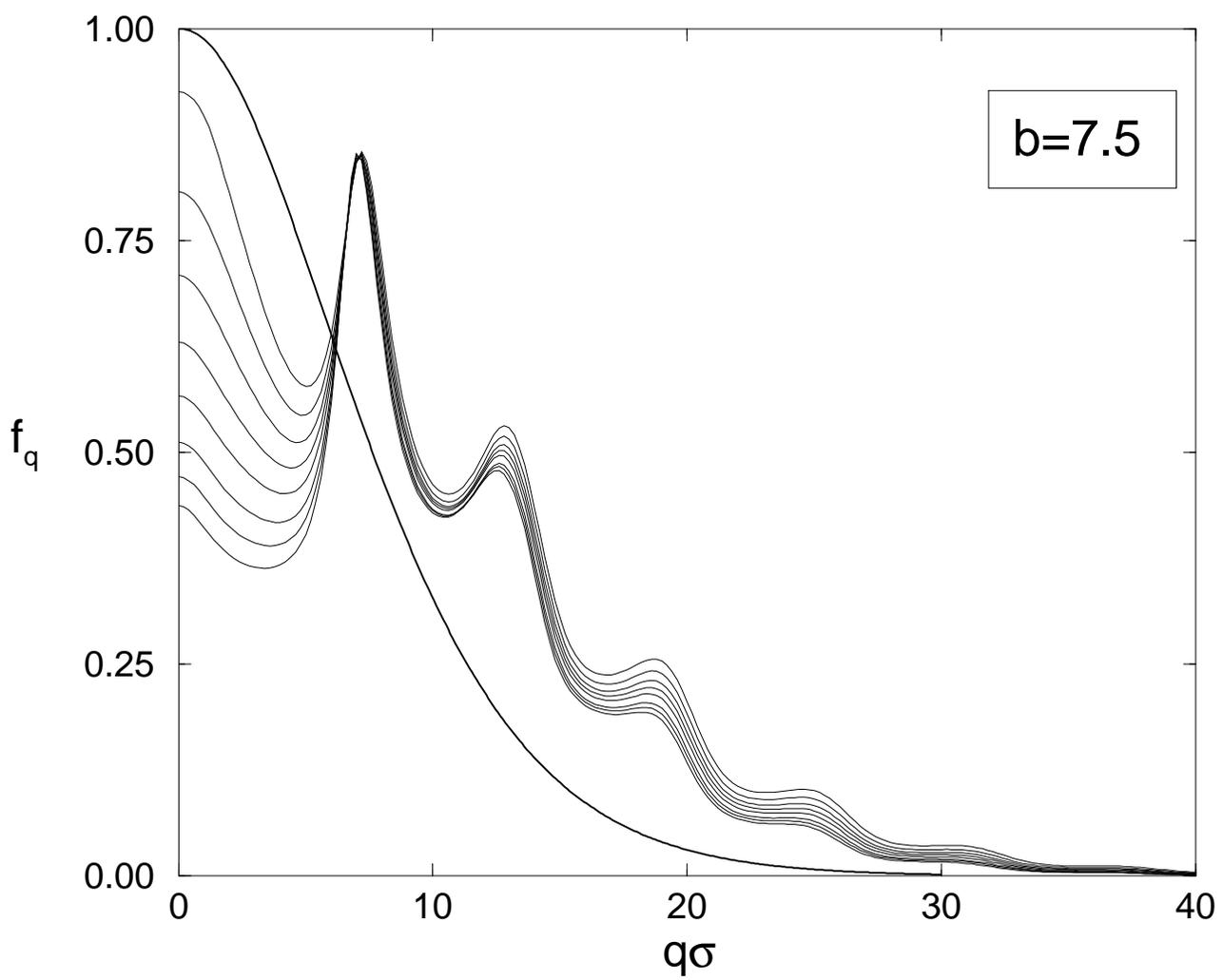
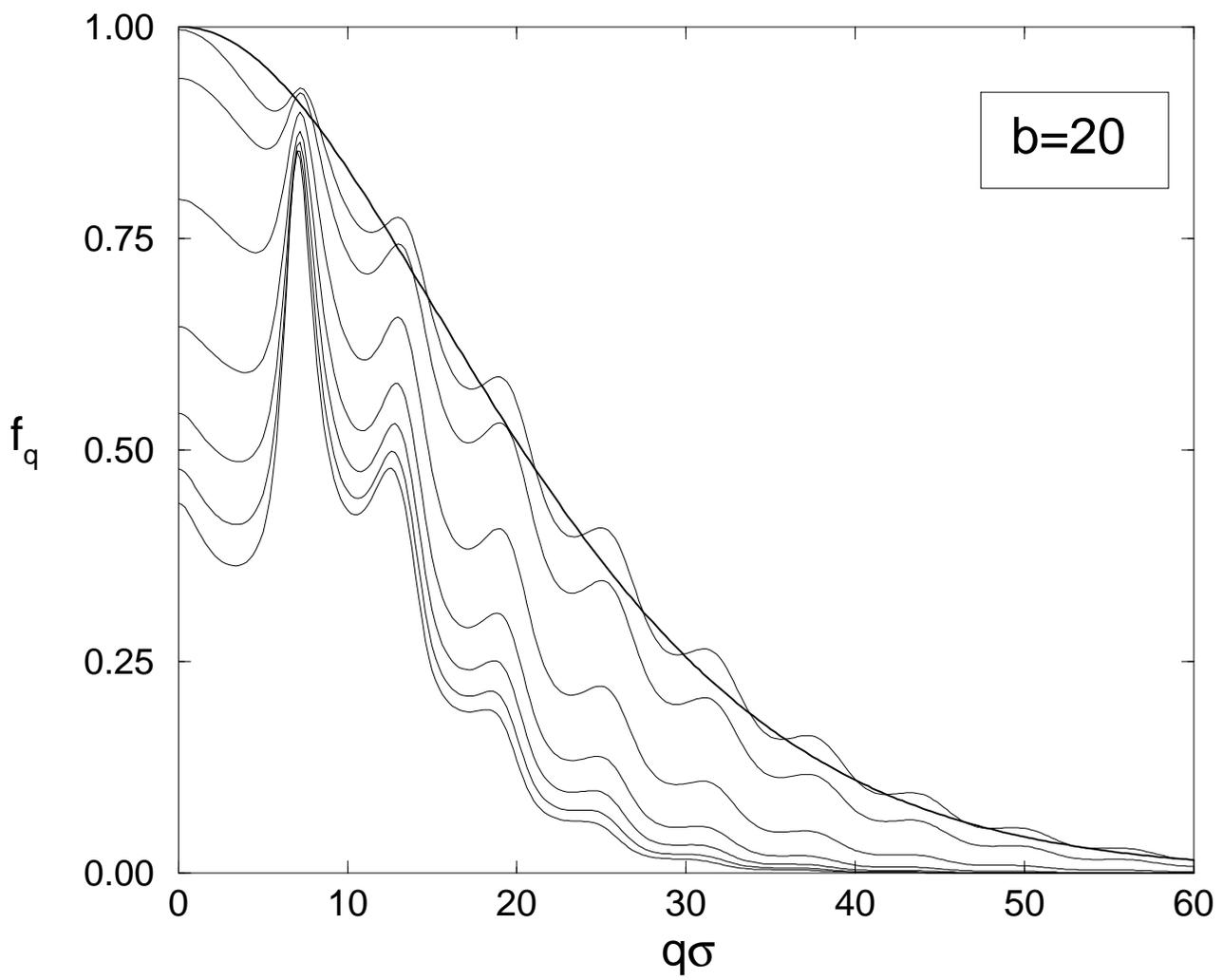
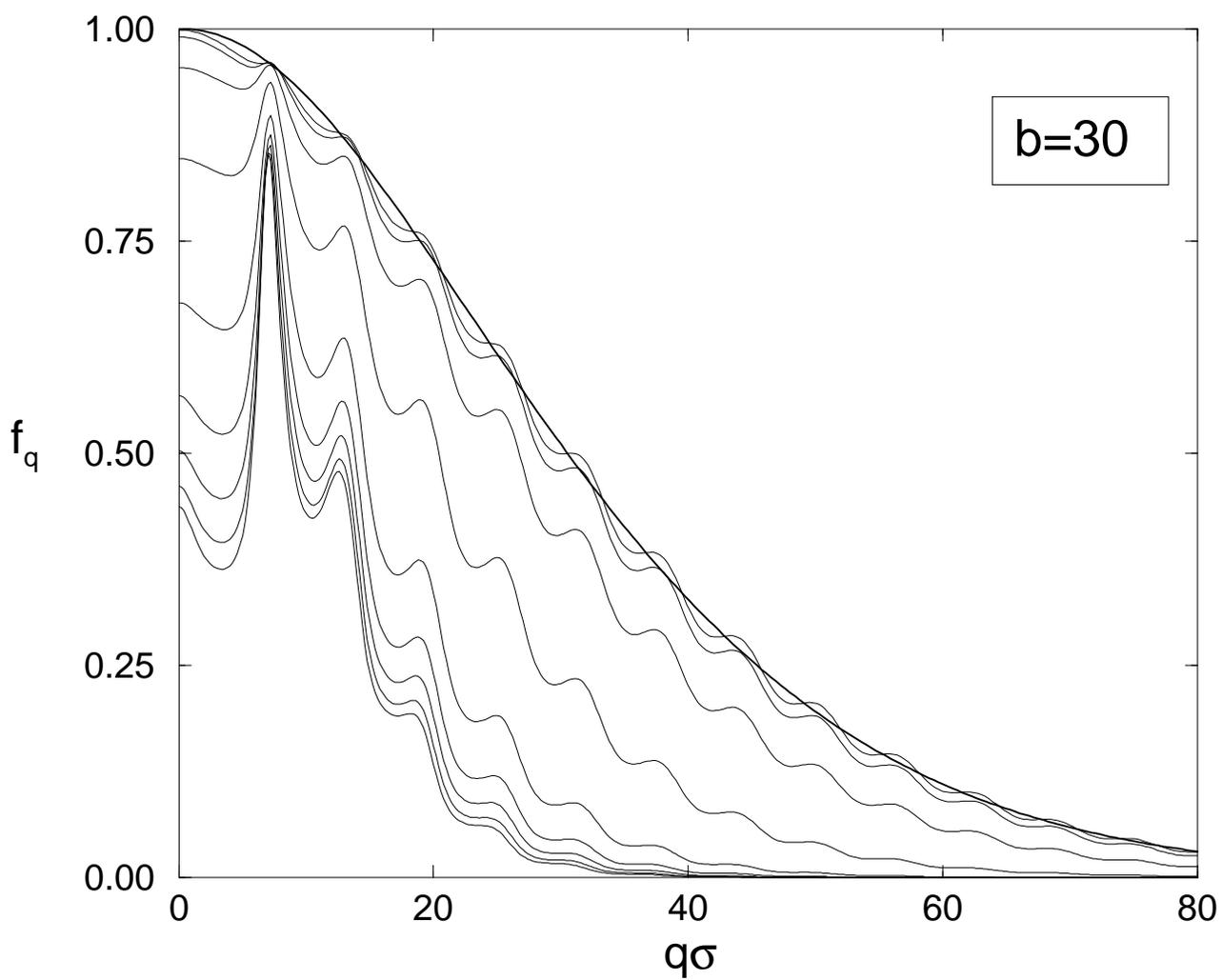
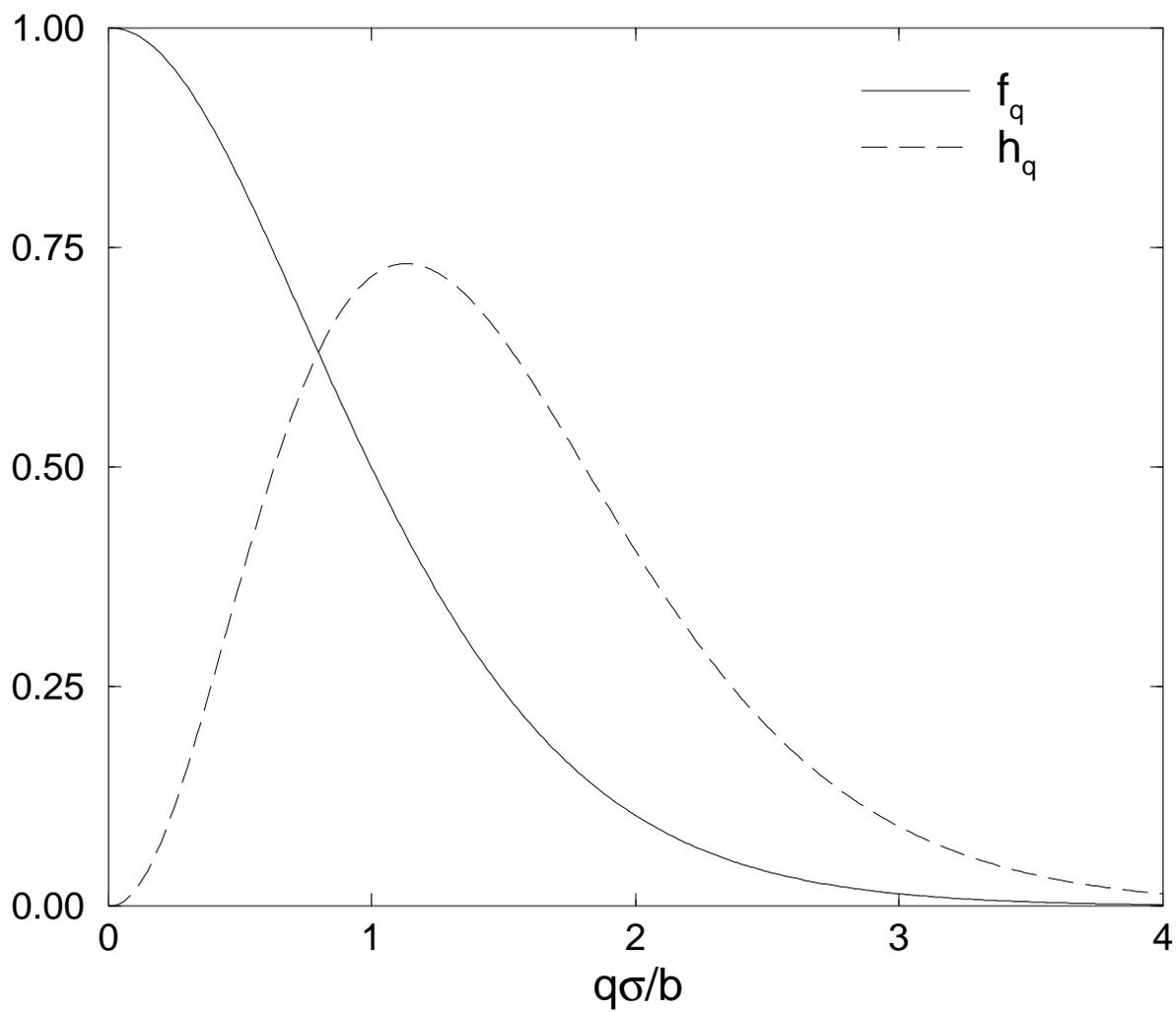
\end{document}